\documentclass{iopart} 
\usepackage{graphicx}
\usepackage[]{epsfig}
\usepackage{iopams}

\newcommand{\rref}[1]{reference~\cite{#1}}

\newcommand{\rsref}[1]{references~\cite{#1}}

\newcommand{\Eq}[1]{Eq.~(\ref{#1})}

\newcommand{\Dinf}{\Delta_{\infty}}
\newcommand{\hinf}{h_{\infty}}

\begin{document}

\title[Sedimentation of Colloidal Gels] {The Role of Solid Friction in the Sedimentation of Strongly Attractive Colloidal Gels}

\author{Jean-Michel Condre, Christian Ligoure, and Luca Cipelletti\footnote[3]{To whom correspondence should be addressed
(lucacip@lcvn.univ-montp2.fr)}}

\address{LCVN UMR 5587, Universit\'{e} Montpellier II and CNRS, P. Bataillon, 34095 Montpellier, France}

\begin{abstract}

We study experimentally and theoretically the sedimentation of gels
made of strongly aggregated colloidal particles, focussing on the
long time behavior, when mechanical equilibrium is asymptotically
reached. The asymptotic gel height is found to vary linearly with
the initial height, a finding in stark contrast with a recent study
on similar gels [Manley \textit{et al.} 2005 \textit{Phys. Rev.
Lett.} \textbf{94} 218302]. We show that the asymptotic compaction
results from the balance between gravity pull, network elasticity,
and solid friction between the gel and the container walls. Based on
these ingredients, we propose a simple model to account for the
dependence of the height loss on the initial height and volume
fraction. As a result of our analysis, we show that the static
friction coefficient between the gel and the container walls
strongly depends on volume fraction: the higher the volume fraction,
the weaker the solid friction. This nonintuitive behavior is
explained using simple scaling arguments.

\end{abstract}


\submitto{JSTAT}


\section{Introduction} \label{sec:intro}

Most colloidal systems tend to aggregate, unless special care is
taken to stabilize them, e.g. by designing particles with surface
charges or a sterically stabilizing polymer layer that prevent them
to approach too closely, where short-ranged, strong Van der Waals or
hydrophobic/hydrophilic attractive interactions prevail. As a result
of aggregation, ramified structures, often with a fractal
morphology, are formed. For high enough particle volume fraction,
$\varphi_0$, and interaction strength, these aggregates form a
space-filling network which is termed a colloidal gel~\cite{Larson}.
The fate of colloidal gels depends crucially on gravity, since the
magnitude of both their elastic modulus and yield stress are
generally comparable to the gravitational stress due to the mismatch
between the density of the particles and that of the solvent in
which they are dispersed.

Because colloidal aggregates and gels are ubiquitous in the food,
drug, cosmetic and oil industry and in biological systems, a large
number of works has been devoted to the sedimentation (or the
creaming) of aggregated colloidal systems. Most investigations have
dealt with``weak'' gels, where the interparticle attraction is
comparable to $k_{\mathrm B}T$, the thermal
energy~\cite{ParkerFood1995,PoonFaradayDiscuss1999,VerhaeghPhysicaA1997,LeePRE2006}.
Experimentally, this is typically the situation encountered in
systems where the attraction between the colloids is induced by the
addition of smaller particles or polymers via the so-called
depletion mechanism~\cite{PoonJPCM2002}. For these systems, various
sedimentation regimes have been observed, ranging from ``creeping''
sedimentation (a linear decrease of the gel height, $h$, with time
$t$) to ``delayed'' sedimentation ($h(t)$ is nearly constant during
an initial latency time, followed by a sudden collapse of the gel
and a final slow compaction of the sediment). Moreover, the
sedimentation of samples with identical composition may drastically
differ according to the container size and shape and in particulary
depending on the initial height of the gel, $h_0$. This behavior can
be rationalized according to whether or not stress can be
transmitted through the sample over distances comparable to the
container size~\cite{StarrsJPCM2002,EvansJPCM2002}. This is the case
for the strongest gels, where the attractive interactions are large
enough to make it likely that connecting paths along the gel network
persist, in spite of the continuous breaking and reforming of
particle bonds due to thermal energy.

The group of C. Allain has extensively studied stronger gels, where
particle bonds are unlikely to break because of thermal
energy~\cite{AllainPRL1995,SenisPRE1997,SenisEPJE2001,DerecPRE2003}.
These gels, however, are still weak enough to experience extensive
restructuring due to the gravitational stress. For the most diluted
samples, the very formation of a gel is prevented, because clusters
grow more slowly than in concentrated samples and they sediment
before getting a chance to form a system-spanning network. By
contrast, Manley and coworkers have studied ``strong'' gels made of
silica particles where neither thermal energy nor gravitational
stress are large enough to cause significant restructuring of the
network~\cite{ManleyPRL2005Sedimentation}. They find that $h(t)$
relaxes exponentially towards an asymptotic height $\hinf$ and
propose that this behavior results from the competition between
network elasticity, gravitational pull, and viscous drag of the
solvent through the gel pores.

In this paper, we study strongly attractive colloidal gels similar
to those of Manley \textit{et al.}, albeit at slightly smaller
volume fractions. Although we observe a similar exponential
evolution of $h(t)$, we find that $\Dinf \equiv h_0-\hinf$ is
proportional to the initial height, a behavior incompatible with
that predicted by the model of \rref{ManleyPRL2005Sedimentation}.
The volume fraction dependence of the sedimentation is also found to
deviate with respect to that postulated
in~\cite{ManleyPRL2005Sedimentation}. We propose a more general
model for the equilibrium behavior of strongly attractive colloidal
gels: in addition to network elasticity and gravitational stress,
our model introduces a term describing the solid friction between
the gel and the cell walls, using a formalism similar to that first
developed by Janssen for granular materials~\cite{Janssen1895}. The
model accounts for all our experimental observations and allows us
to estimate the static friction coefficient, $\mu_s$, between the
gel and the walls. Quite surprisingly, we find that $\mu_s$
\emph{decreases} with increasing volume fraction. We propose a
simple scaling argument to show that this counterintuitive behavior
is due to the fractal nature of the aggregates forming the gel.

\section{Materials and methods}
\label{sec:matmet}

We use aqueous dispersions of charge-stabilized silica spheres
(Ludox TM-50) with radius $a\approx 11$ nm. Aggregation is induced
by mixing equal volumes of a particle suspension and a NaCl solution
to a final salt concentration of 2M and final particle volume
fractions $\varphi_0$ ranging from $2.5\times10^{-3}$ to $10^{-2}$.
At this ionic strength, electrostatic repulsion becomes negligible
and the particles experience short-ranged, strong van der Waals
interactions, leading to the formation of aggregates in the
diffusion limited cluster aggregation (DLCA) regime
\cite{WeitzPRL1984}. The bond energy is of the order of several tens
of $k_{\rm B}T$, making thermally activated rearrangements extremely
unlikely. The aggregates have a fractal dimension $d_{f} \approx
1.9$; during aggregation their typical size increases until they
form a space-filling network. When this gelled structure is formed,
the aggregates have a size distribution peaked about an average
cluster size $\xi \approx
a\varphi_{0}^{1/(d_{f}-3)}$~\cite{CarpinetiPRL1992}. The gelation
time can be calculated within the DLCA model: for our gels it is
less than a few seconds~\cite{LinJPCM1990}.

The gels are viscoelastic materials: oscillatory strain measurements
show that over a large range of frequencies (typically
$10^{-2}~\mathrm{rad/sec}~< \omega < 10~\mathrm{rad/sec}$) the
storage modulus $G'$ is nearly frequent-independent and about one
order of magnitude larger than the loss modulus
$G"$~\cite{GislerPRL1999,ManleyPRL2005SilicaMicrog}. Due to the
fractal morphology of the gels, $G'$ increases very strongly with
volume fraction: $G' \propto \varphi_{0}^{\nu}$ where $\nu \approx
3.6-4.0$
\cite{BuscallJNNMF1987,KrallPRL1998,GislerPRL1999,ManleyPRL2005SilicaMicrog}.
Moreover, the gels are able to sustain large shear strains up to
about $10\%$ before the network is significantly altered and linear
elasticity fails~\cite{GislerPRL1999}. Applying larger strains leads
initially to strain hardening, but eventually results in the
disruption of the gels, although gel breaking may not occur
instantaneously~\cite{GislerPRL1999,ManleyPRL2005Sedimentation}.

The particle density is $2.35~\mathrm{g/cm^{3}}$, while that of the
solvent is $1.15~\mathrm{g/cm^{3}}$ (due to the high salt
concentration). The density mismatch, $\Delta
\rho=1.2~\mathrm{g/cm^{3}}$, is large enough to make the gels
sensitive to gravitational stress, leading to a slow compaction.
Experiments are performed by directly preparing the gels in glass
cylindrical cells of diameter $D=1.7~\mathrm{cm}$ and letting them
settle unperturbed. Note that contrary to the protocol of
\rref{SenisPRE1997} no stirring is applied after the initial mixing
of the particle and salt solutions. The gels are imaged with a CCD
camera to capture the settling kinetics. We check that no cracks
appear in the bulk of the gels. As observed in other
experiments~\cite{PoonFaradayDiscuss1999,StarrsJPCM2002}, at the
earliest stages of sedimentation the meniscus at the top of the
suspension is emptied from the particles. Once only solvent is left
in the meniscus, the gel starts settling uniformly throughout the
cross section of the cell. A reasonably flat, sharp interface is
observed between the highly turbid gel and the clear supernatant,
allowing us to measure the time evolution of the height of the gel,
$h(t)$, with an accuracy of about $5\%$. We define $t=0$ as the time
when the meniscus is emptied and the uniform settling starts. We
study gels at six volume fractions: $\varphi_{0} =
2.5\times10^{-3}$, $3\times10^{-3}$, $5\times10^{-3}$,
$6\times10^{-3}$, $7.5\times10^{-3}$, and $10^{-2}$. For
$\varphi_{0} < 2.5\times10^{-3}$ the gels often develop cracks and
the experiments are less reproducible, while for $\varphi_{0} >
10^{-2}$ the gels hardly settle, making a precise determination of
$h(t)$ impossible with our apparatus. For each $\varphi_{0}$, we
vary the initial height of the gel, $h_{0}$, in the range
$0.5~\mathrm{cm}<h_{0}<5~\mathrm{cm}$.

\section{Experimental results}
\label{sec:results}

\begin{figure}
\begin{center}
\psfig{file=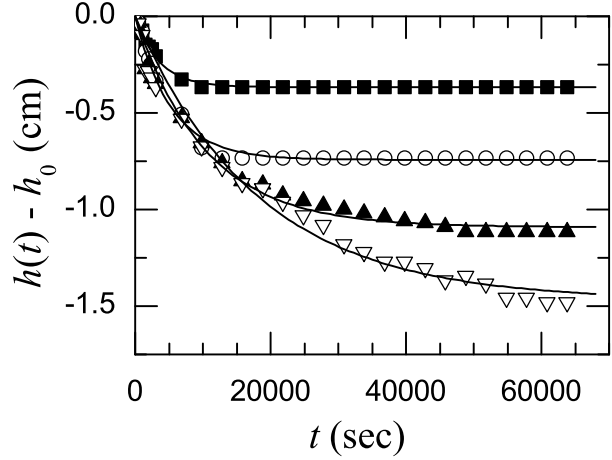,width=8.cm}
\end{center}
\caption{\label{Fig:deltahvstime} Time dependence of the settling
of a gel with $\varphi_0 = 0.3\times 10^{-2}$. From top to bottom,
the initial height of the gel, $h_0$, is 1.23, 2.45, 3.68, and 4.9
cm. The lines are fits of \Eq{Eq:DeltaHVsT} to the data.}
\end{figure}

Figure~\ref{Fig:deltahvstime} shows the evolution of the gel height
measured for $\varphi_0 = 0.3\times 10^{-2}$ and various initial
heights $h_0$. The time evolution of $h(t)$ is well fitted by a
simple exponential relaxation:
\begin{equation}
\label{Eq:DeltaHVsT} h_{0}-h(t)=\Dinf\left [ 1-\exp(-t/\tau)\right
]
\end{equation}
where $\Dinf=| h_{\infty} -h_{0}|$ denotes the asymptotic settling
and $h_{\infty} = h(t\rightarrow \infty)$. A similar behavior has
been reported recently for the ``strong'' gels of
reference~\cite{ManleyPRL2005Sedimentation}. Both the asymptotic
settling and the characteristic time $\tau$ grow with the initial
height of the gel; for the tallest gels ($h_0 = 4.9$ cm), $\tau$ is
as large as a few hours. The same exponential relaxation,
\Eq{Eq:DeltaHVsT}, describes the height evolution of all the gels
that we have studied, except for some samples where fractures
appeared during the settling. For fractured gels, $h(t)$ deviates
from an exponential relaxation and the asymptotic settling is larger
than that expected for pristine samples. In the following we analyze
only data for non-fractured gels.

\begin{figure}
\begin{center}
\includegraphics[scale=1]{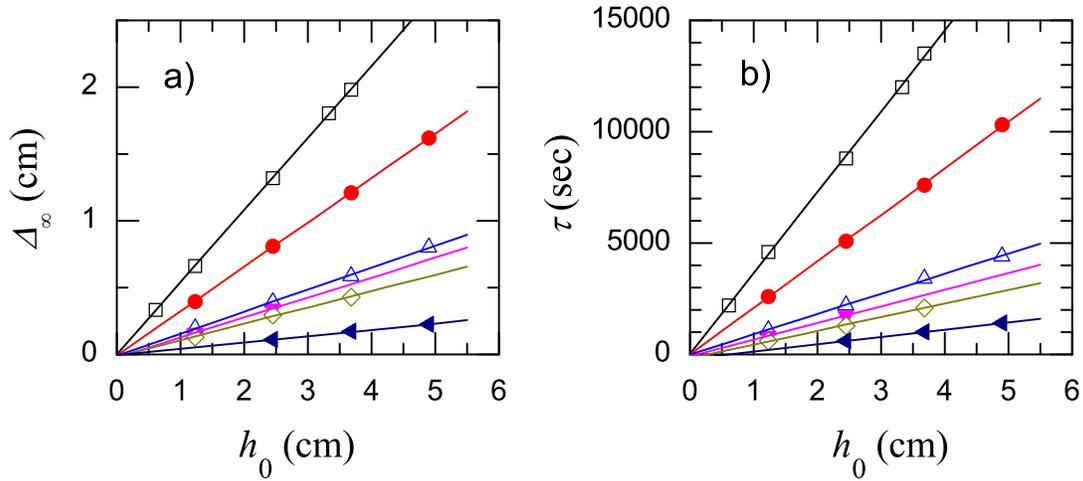}
\end{center}
\caption{\label{Fig:deltahvsh0} Asymptotic settling of the gel
height $\Dinf$ (a), and characteristic time $\tau$ for the
sedimentation (b) as a function of the initial height $h_0$. From
top to bottom, $\varphi_{0} = 2.5\times10^{-3}$, $3\times10^{-3}$,
$5\times10^{-3}$, $6\times10^{-3}$, $7.5\times10^{-3}$, and
$1\times10^{-2}$. Both quantities vary linearly with $h_0$, as
shown by the fits (straight lines).}
\end{figure}

We plot in Fig.~\ref{Fig:deltahvsh0} the $h_0$ dependence of the
asymptotic settling (a) and of the characteristic settling time (b).
Remarkably, both $\Dinf$ and $\tau$ vary linearly with the initial
height of the gel, as shown by the lines that are fits of form
$\Dinf = \alpha h_0 + \alpha_1$ and $\tau = \beta h_0 + \beta_1$,
respectively. As one can see in Figure~\ref{Fig:deltahvsh0}, both
$\alpha_1$ and $\beta_1$ are almost zero, making the fits
indistinguishable from straight lines through the origin. In fact,
we find that $\beta_1 = 0$ is compatible with experimental
uncertainty, while $\alpha_1$ is slightly but consistently smaller
than 0, a feature that will be commented further below.

\begin{figure}
\begin{center}
\includegraphics[scale=1]{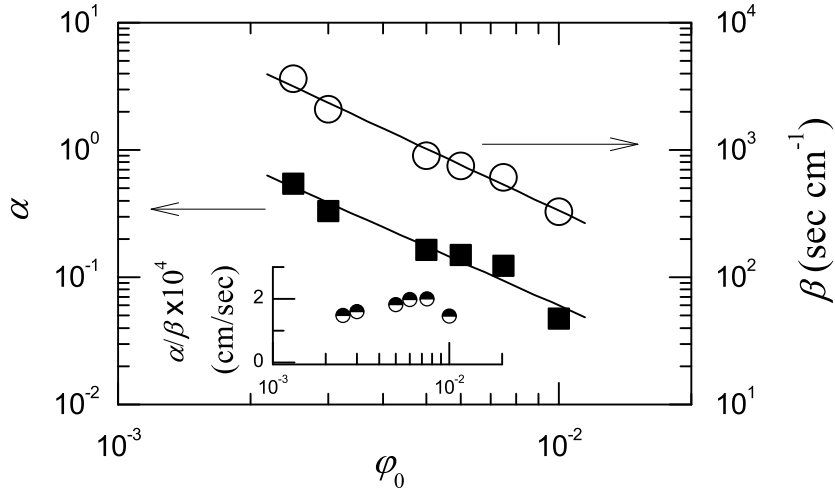}
\end{center}
\caption{\label{Fig:alphabeta} Left axis, solid squares: $\varphi_0$
dependence of the proportionality coefficient $\alpha$ between
$\Dinf$ and $h_0$ extracted from the linear fits of
Fig.~\ref{Fig:deltahvsh0}a). The line is a power law fit yielding an
exponent $-1.54 \pm 0.17$. Right axis, open circles: proportionality
coefficient $\beta$ between $\tau$ and $h_0$ extracted from the
linear fits of Fig.~\ref{Fig:deltahvsh0}b). The line is a power law
fit yielding an exponent $-1.62 \pm 0.11$. Inset: ratio
$\alpha/\beta$ \textit{vs} $\varphi_0$. Note the weak variation of
$\alpha/\beta$ with volume fraction, indicating that the initial
sedimentation velocity is almost constant, as discussed in the
text.}
\end{figure}

For a fixed cell diameter and material, the prefactors $\alpha$ and
$\beta$ depend solely on volume fraction. Their $\varphi_0$
dependence is shown in Fig.~\ref{Fig:alphabeta} in a log-log scale.
The data can be fitted reasonably well by power laws: $\alpha \sim
\varphi^{u}_{0}$ and $\beta \sim \varphi^{v}_{0}$, with $u = -1.54
\pm 0.17$ and $v = -1.62 \pm 0.11$. Quite intriguingly, $u \approx
v$: this indicates that for $t\rightarrow 0$ the settling velocity
of the gels is nearly independent of both their height and volume
fraction. Indeed, for $t<<\tau$ \Eq{Eq:DeltaHVsT} yields $|dh/dt|
\approx \Dinf/\tau = \alpha/\beta$, which varies weakly between
$1.45 \times10^{-4}$ and $2 \times10^{-4}$ cm/sec as shown in the
inset of Fig.~\ref{Fig:alphabeta}.

In order to understand the behavior of our gels, we start by
comparing the results presented in
Figures~\ref{Eq:DeltaHVsT}--\ref{Fig:alphabeta} with previous works.
C. Allain and coworkers have measured the quantity $\Omega =
1-\Dinf/h_0 = 1-\alpha$ and found that in the so-called
``friction-dominated'' regime $\Omega$ is independent of the initial
height $h_0$~\cite{AllainPRL1995,SenisPRE1997}, in agreement with
our observation $\Dinf \approxeq \alpha h_0$. However, they find
$\Omega \propto \varphi_0$, a scaling that is incompatible with the
power law $\alpha \sim \varphi^{-1.62}$ shown in
Figure~\ref{Fig:alphabeta}, as we also check by plotting directly
$\Omega$ \textit{vs} $\varphi_0$ (data not shown). This discrepancy
suggests that the physical mechanism leading to the settling of our
gels is different from that of \rsref{AllainPRL1995,SenisPRE1997}.
Indeed, the gels studied by Allain's group experience extensive
restructuring and fractures during
sedimentation~\cite{SenisEPJE2001,DerecPRE2003}, in contrast to our
gels. During sedimentation, the gel does not retain its integrity
and the scaling $\Omega \propto \varphi_0$ results from the
formation of a sediment composed of randomly packed aggregates
issued from the gravity-induced breaking of the gel.

An exponential relaxation similar to that shown in
Figure~\ref{Eq:DeltaHVsT} has been reported for gels made of silica
particles by Manley \textit{et.~al}, who studied the regime where no
fractures are observed~\cite{ManleyPRL2005Sedimentation}, similarly
to our experiments. This sedimentation kinetics is interpreted as
the result of the interplay between gravitational pull, network
elasticity, and viscous drag of the solvent through the network
pores. However, our results differ distinctly from those of
\rref{ManleyPRL2005Sedimentation} in that we find i) $\Dinf \propto
h_0$ rather than $\Dinf \propto h_{0}^2$ as predicted by the model
of~\cite{ManleyPRL2005Sedimentation}, ii) $\tau \propto h_0$, as
opposed to $\tau \propto h_{0}^2$, iii) $\tau \propto
\varphi_{0}^{-1.62}$, rather than $\tau \propto (1-\varphi_{0})$;
iv) $dh/dt|_{t=0} \approx \mathrm{constant}$, while Manley and
coworkers find $dh/dt|_{t=0} \propto \varphi^{(1-d_{\rm
f})/(3-d_{\rm f})} \approx \varphi^{-0.8}$. Therefore, it is clear
that in the case of our gels a fundamental physical ingredient is
missing in the model of ~\rref{ManleyPRL2005Sedimentation}.

\section{Model and comparison with the experiments}
\label{sec:model}

We propose that, in addition to the mechanisms listed above, solid
friction between the gel and the cell walls also plays a crucial
role in the settling of our samples. A hint of the importance of
solid friction is provided by the observation that both $\Dinf$ and
$\tau$ are modified by using plexiglas cells rather than glass
cells~\cite{noteplexi}. Moreover, the importance of solid friction
is consistent with the arguments of \rref{EvansJPCM2002}, where it
was argued that, for gels with strong interparticle interactions,
the characteristic length over which stress can be propagated
exceeds the container dimensions. Accordingly, ``strong'' gels such
as those studied here should feel the influence of the cell walls.
Solid friction is particularly appealing since it provides a simple
physical mechanism to explain the linear scaling of $\Dinf$ with
initial height $h_0$, as shown by the following qualitative
argument. Solid friction screens the gravitational load acting on
any given cross sectional plane $\Sigma$ in the gel: because part of
the weight of the gel column laying above $\Sigma$ is sustained by
the walls, it is as if in $\Sigma$ the gel experienced only the
weight of a column segment of effective height $L$, shorter than the
whole column. One can then imagine to divide a gel of height $h_0$
into segments of height $L$: under gravitational stress, each
segment is compressed by the same amount, regardless of the number
of segments that compose the gel. Therefore, the total compression
is simply proportional to the number of segments contained in the
gel, i.e. $\Dinf \propto h_0/L$.

To make this argument more quantitative, we build a model for the
settling of the gels, assuming that they are elastic media that
compresses under their own weight, without any restructuring, and
subject to solid friction against the cell walls. The rate of
compression is limited both by the viscous friction of the fluid
through the network, which can be viewed as a porous
medium~\cite{SenisEPJE2001,ManleyPRL2005Sedimentation}, and by solid
friction. In the following, we will focus on the behavior for $t>>
\tau$, when mechanical equilibrium is asymptotically reached and
gravitational stress is counterbalanced by both the gel elasticity
and solid friction; the kinetics of sedimentation will be addressed
in future work. We choose the $z$ axis along the vertical direction,
with $z=0$ the bottom of the gel and $z=h_0$ its top at time $t=0$.
Due to gravity, the gel is submitted to a uniaxial compression
$p(z)=-\sigma _{zz}$ where $\sigma _{zz}$ is the vertical stress.

We first recall the behavior in the absence of wall friction.
Newton's law for a gel slice of thickness $dz$ at height $z$ yields

\begin{equation}
-\Delta \rho g \varphi(z)\frac{\pi D^{2}}{4} dz -\frac{\partial
p}{\partial z} \frac{\pi D^{2}}{4}dz =0 \, , \label{Eq:barometric}
\end{equation}
where  $\varphi(z)$ is the local volume fraction, $D$ the cell
diameter and $g$ the acceleration of gravity. The first term
describes the gravity pull exerted on the slice, while the second
term accounts for the elastic response of the material. The
conservation of the total number of particles imposes the additional
condition
\begin{equation}
\int_{0}^{h_{\infty}}{\varphi(z)dz} = \int_{0}^{h_{0}}{\varphi_0dz
= }h_{0} \varphi_0\, , \label{Eq:conservation}
\end{equation}
while the boundary condition is
\begin{equation}
\varphi(h_{\infty}) = \varphi_0\, , \label{Eq:boundarycondition}
\end{equation}
since the top layer of the gel is uncompressed. If the deformation
is not too large, the pressure and the volume fraction are related
by the equation of state of an elastic medium in the linear regime:
$p(z)= C\frac{\varphi(z)-\varphi_{0}}{\varphi_{0}}$, where the
uniaxial compression modulus $C$ is a linear combination of $B$, the
bulk modulus, and the shear modulus: $C=B+ 4G/3$.
Equation~(\ref{Eq:barometric}) can be solved for $\varphi(z)$; by
imposing the boundary condition, Eq.~(\ref{Eq:boundarycondition}),
one obtains the asymptotic profile:
\begin{equation}
\varphi(z) = \varphi_0 \exp \left [(\hinf-z)/\lambda \right ] \, ,
\label{Eq:profilebarometric}
\end{equation}
where we have introduced $\lambda=C/(\Delta \rho g \varphi_{0})$,
the characteristic length over which an elastic material is deformed
due to gravity (the stiffer the gel, or the weaker the density
mismatch, the larger $\lambda$). By requiring that
Eq.~(\ref{Eq:profilebarometric}) satisfies particle conservation
(Eq.~\ref{Eq:conservation}), one finds $\Dinf = h_0 -
\lambda\ln(1+h_0/\lambda)$. In the limit $h_0 << \lambda$, this
relation simplifies to $\Dinf = h_{0}^2/(2\lambda) +
\lambda\mathcal{O}(h_{0}/\lambda)^3$, whose leading term coincides
with the expression given in \rref{ManleyPRL2005Sedimentation}. We
estimate $G \approx 7~\mathrm{Pa}$ for our gel at $\varphi_0 = 1
\times 10^{-2}$ by extrapolating the data of
ref.~\cite{ManleyPRL2005SilicaMicrog}. By assuming that the order of
magnitude of $C$ is the same as that of the shear modulus plateau,
one estimates $\lambda \approx 13~\mathrm{cm}$. [Using the full
model that takes into account solid friction, we will obtain
$\lambda$ directly from the experimental data, finding a result
consistent with this estimate, see Fig.~\ref{Fig:ellelambdavsphi}
below]. Hence, $h_0 < \lambda$ in all our experiments (indeed, in
most cases $h_0 << \lambda$) and a quadratic scaling of $\Dinf$ with
$h_0$ should be observed. By contrast, as recalled above, we find
$\Dinf \propto h_0$, indicating that gravity and elasticity alone
are not sufficient to correctly describe the behavior of our
samples.

Solid friction has been first modeled for a container filled with a
granular material by Janssen in 1895~\cite{Janssen1895}. In the
Janssen model, the pressure $p(z)$ is partially redirected in the
horizontal direction leading to an horizontal stress $
\sigma_{rr}(z)= K p(z)$ with $K=\frac{B-2G/3}{B+4G/3} $ the
redirection coefficient. When mechanical equilibrium is reached, the
effective static friction of a slice of material of thickness $dz$
confined in a cylinder of diameter $D$ is oriented upward and has
magnitude $\pi \mu_{s} D Kp(z)dz$, where $\mu_{s}$ is the static
friction coefficient between the grains and the container wall. The
Janssen model predicts that, when moving from the top of the grain
column downward, the pressure increases reaching exponentially a
saturation value with a characteristic screening length $L =
D/(4\mu_{s} K)$. This is very different from the case of a liquid
where the pressure increases linearly with depth: the saturation is
due to the fact that the cylinder wall supports part of the weight
of grains: the contribution to the vertical stress on a horizontal
plane $\Sigma$ due to the material at a height $\Delta z$ above
$\Sigma$ decreases as $\exp(-\Delta z / L)$. Hence, the larger the
contribution of solid friction, the smaller $L$. The universality of
the Janssen scaling for the stress saturation curve has been
recently shown by Ovarlez \textit{et al.}~\cite{OvarlezPRE2003}.
Moreover, the Janssen model has been generalized to dynamical
situations~\cite{BoutreuxPRE1997}.

We propose to describe the solid friction between the gels and the
cell walls using Janssen's formalism. At equilibrium,
Eq.~(\ref{Eq:barometric}) becomes
\begin{equation}
-\Delta \rho g \varphi(z)\frac{\pi D^{2}}{4} dz -\frac{\partial
p}{\partial z} \frac{\pi D^{2}}{4} dz + \pi \mu_{s} D Kp(z)dz =
0\, . \label{Eq:fullmodel}
\end{equation}
It is convenient to solve Eq.~(\ref{Eq:fullmodel}) for the local
variation in volume fraction, $\Delta \varphi(z) =
\varphi(z)-\varphi_{0}$, rather than directly for $\varphi(z)$.
Assuming again linear elasticity and using the boundary condition
(\ref{Eq:boundarycondition}), after some standard manipulations
one finds the equilibrium concentration profile:
\begin{equation}
\Delta \varphi(z)=\varphi_{0}\frac{\ell}{\lambda} \left[1-\exp
\left( \frac{z-h_{\infty}}{\ell}\right) \right]
\label{Eq:profilefullmodel}
\end{equation}
where we have defined a third length $\ell$ related to the elastic
and solid friction characteristic lengths introduced above by
$1/\ell=1/L-1/\lambda$. Note that there are no \textit{a priori}
restrictions on the sign of $\ell$; $\ell = 0$ is also possible, if
the elastic and friction characteristic lengths are equal. We will
show in the following that for our gels $0 < \ell < h_0$: in this
case Eq.~(\ref{Eq:profilefullmodel}) predicts the concentration
profile to be essentially constant, except for a segment of length a
few $\ell$'s at the top of the gel. In general, we expect both
$\lambda$ and $L$ to depend on the gel volume fraction $\varphi_0$;
additionally, $L$ may contain any deviations with respect to the
usual Jannsen's law for ordinary solid materials, e.g. due to the
fractal morphology of the gels. For $L \rightarrow \infty$, solid
friction becomes negligible and Eq.~(\ref{Eq:profilefullmodel})
reduces to the exponential profile predicted by
Eq.~(\ref{Eq:profilebarometric}). By imposing particle conservation,
$\Dinf$ is found to satisfy
\begin{equation}
\Dinf=\frac{\ell}{\lambda}\left\{ h_{\infty}-\ell
\left[1-\exp(-h_{\infty}/\ell)\right ]\right\} \, .
\label{Eq:Deltahmodel}
\end{equation}
Because $h_{\infty} = h_0 - \Dinf$, this expression is an implicit
equation for $\Dinf$ when $\lambda$ and $L$ are known and the
initial height is assigned. By contrast,
Eq.~(\ref{Eq:Deltahmodel}) may be used directly to fit the
experimental data, which can be easily plotted as $\Dinf$
\textit{vs} $h_{\infty}$, rather than \textit{vs} $h_{0}$ as in
Fig.~\ref{Fig:deltahvsh0}.

\begin{figure}
\begin{center}
\includegraphics[scale=1.5]{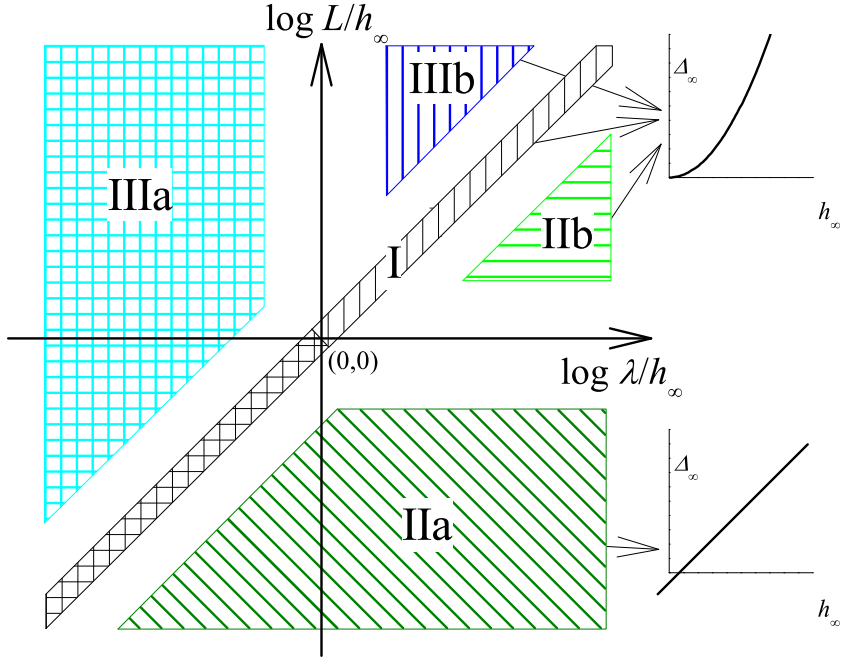}
\end{center}
\caption{\label{Fig:statediagram} ``State diagram'' illustrating the
location of the various asymptotic settling regimes discussed in the
text in the ($\log\lambda/\hinf$, $\log L/\hinf$) plane. The
corresponding approximate behavior of $\Dinf$ as a function of
$\hinf$ is shown on the right. The $\lambda < \hinf$ portion of zone
I and zone IIIa should be disregarded, since they correspond to a
large deformation regime that is incompatible with the assumption of
linear elasticity.}
\end{figure}

Various settling regimes are possible, depending on the relative
values of $\lambda$, $L$, and $h_0$ (or, equivalently, $\hinf$).
Three limiting cases are identified, according to the behavior of
the argument of the exponential term in the r.h.s of
Eq.~(\ref{Eq:Deltahmodel}): i) $\hinf / \ell \rightarrow 0$, for
which $\Dinf \approx \hinf^2/(2\lambda)$; ii) $\hinf / \ell
\rightarrow \infty$, for which $\Dinf \approx \hinf\ell/\lambda -
\ell^2/\lambda$; iii) $\hinf / \ell \rightarrow -\infty$, for which
$\Dinf \approx (\lambda^2/\ell)\exp(\hinf/|\ell|)$. The assumption
of linear elasticity that was made in deriving
Eqs.~(\ref{Eq:profilefullmodel}) and~(\ref{Eq:Deltahmodel}) poses an
additional constraint on the solutions for $\Dinf$, since linear
elasticity only applies to small deformations $\Dinf \lesssim
\hinf$. We recapitulate the different settling regimes in
Fig.~\ref{Fig:statediagram}, which shows schematically a ``state
diagram'' of the solutions to Eq.~(\ref{Eq:Deltahmodel}), using as
``state variables'' $\lambda/\hinf$ and $L/\hinf$ and classifying
the various zones according to the value of $L$ relative to
$\lambda$. Zone I corresponds to comparable friction and elasticity
characteristic lengths ($|\lambda-L| <<\lambda L / \hinf$). In zones
IIa and IIb the elastic length scale dominates over the friction
length scale ($L/\hinf << 1$ and $L/\hinf << \lambda/\hinf$ for IIa
and $1 << L/\hinf << \lambda/\hinf $ for IIb). The reverse applies
to zones IIIa, where $\lambda/\hinf << 1$ and $\lambda/\hinf <<
L/\hinf$, and IIIb, for which $1 << \lambda/\hinf << L/\hinf $. In
summary, for the $L/\hinf
> 1$ portion of zone I and for zones IIb and IIIb one finds that a quadratic law
applies: $\Dinf \approx \hinf^2/(2\lambda) \approx
h_{0}^2/(2\lambda)$. Note that this is the same expression as
derived in the absence of solid friction. Hence, in these zones and
to leading order solid friction has no influence on the asymptotic
settling. At first sight, this result is somehow surprising for zone
IIb, where $L < \lambda$ and one could have guessed that solid
friction would dominate over elasticity. However, our analysis shows
that solid friction is negligible as long as $L$ is much larger than
the asymptotic height of the gel, regardless of the characteristic
length for elasticity. In zone IIa, a linear growth is found:
\begin{equation} \Dinf\approx
\frac{\ell}{\lambda}\hinf - \frac{\ell^2}{ \lambda} \, ,
\label{Eq:DeltaapproxIIa}
\end{equation}
or, equivalently,
\begin{equation}
\Dinf\approx \frac{L}{\lambda}h_0 - \frac{L^2}{ \lambda-L} \, .
\label{Eq:DeltaapproxIIb}
\end{equation}
As one can easily check, the second term of the r.h.s. of
Eq.~(\ref{Eq:DeltaapproxIIb}) is negative and its magnitude is
small compared to $h_0L/\lambda$, yielding a nearly linear
dependence of $\Dinf$ on $h_0$, with a slightly negative $h_0 = 0$
intercept. This is precisely the behavior observed experimentally
for our gels (see Fig.~\ref{Fig:deltahvsh0} and related
discussion), strongly suggesting that our samples belong to zone
IIa, where both elasticity and solid friction are present, but the
contribution of the latter dominates. Finally, the $L/\hinf < 1$
portion of zone I and zone IIIa correspond to large deformations
$\Dinf > \hinf$ that are incompatible with the assumption of
linear elasticity and thus should be disregarded.

\begin{figure}
\begin{center}
\includegraphics[scale=1]{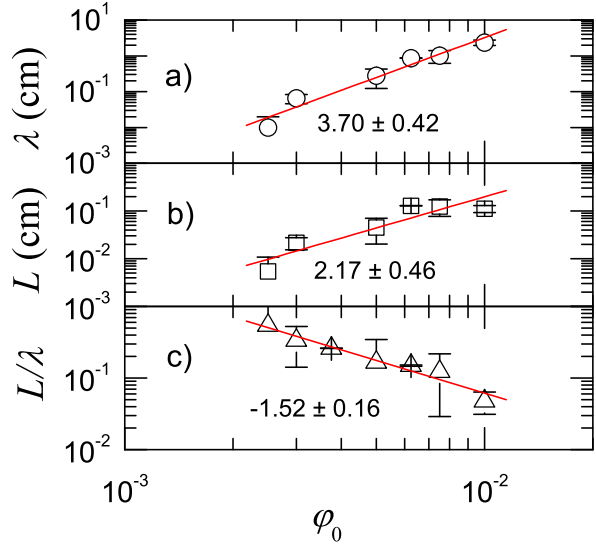}
\end{center}
\caption{\label{Fig:ellelambdavsphi} Volume fraction dependence of
a) the elastic characteristic length $\lambda$, b) the solid
friction characteristic length $L$, c) the ratio $L/\lambda$. The
lines are power law fits to the data, whose exponent is shown in the
panels. The values shown here confirm that our gels belong to zone
IIa of Fig.~\ref{Fig:statediagram}, as suggested by the linear
dependence of $\Dinf$ \textit{vs} $h_0$. }
\end{figure}

To further test our model, we extract the two characteristic lengths
$L$ and $\lambda$ from the height dependence of $\Dinf$ and analyze
both their $\varphi_0$ dependence and absolute
magnitude~\cite{notefit}. Their volume fraction dependence is shown
in Fig.~\ref{Fig:ellelambdavsphi}, together with their ratio. The
elastic length $\lambda$ increases as $\varphi_{0}^{3.7 \pm 0.4}$,
in excellent agreement with the scaling reported for the shear
modulus of similar colloidal gels
\cite{BuscallJNNMF1987,KrallPRL1998,GislerPRL1999,ManleyPRL2005SilicaMicrog}.
Moreover, for $\varphi_0 = 10^{-2}$ we find $\lambda = 2.35$ cm,
from which we calculate $C = 2.8~\mathrm{Pa}$. This is of the same
order of magnitude of, albeit somehow smaller than, the elastic
shear modulus $G \approx 7~\mathrm{Pa}$ that we estimate by
extrapolating the data of ref.~\cite{ManleyPRL2005SilicaMicrog} to
$\varphi_0 = 10^{-2}$. The solid friction length scale $L$ is shown
in Fig.~\ref{Fig:ellelambdavsphi}b). If one assumes that the
relation $L = D/(4\mu_sK)$ derived for granular materials by Jennsen
may apply to the gels and furthermore uses $\mu_s \approx 0.6$ and
$K \lesssim 1$ as found typically for macroscopic materials, one
finds $L \gtrsim 0.7$ cm, not much larger than $L \approx 0.1$ cm,
the experimental value for the largest $\varphi_0$. Remarkably,
however, we find that $L$ decreases with decreasing volume fraction,
strongly suggesting that the fractal nature of the gel has to be
taken into account to correctly describe solid friction. Note that
the variation of $L$ with $\varphi_0$ is somehow counterintuitive,
since larger values of $L$ correspond to lesser solid friction:
hence, the higher the volume fraction, the smaller the solid
friction the gel experiences. We will come back to this surprising
behavior in the following. The ratio $L/\lambda$ is found to
decrease with $\varphi_0$, a consequence of the strong increase of
the elastic length scale with volume fraction. As a consistency test
of the theory developed above, we check that the magnitude of
$\lambda$ and $L$ are compatible with the assumption that our gels
belong to zone IIa of the state diagram shown in
Fig.~\ref{Fig:statediagram} ($L << \lambda$ and $L << \hinf$), as
suggested by the linear dependence of $\Dinf$ \textit{vs} $h_0$.
Indeed, we find that for all samples $L/\lambda < 0.54$ and $L/\hinf
< 0.12$. Moreover, the small deformation requirement is reasonably
well fulfilled, since $\Dinf < 0.33 h_0$ for all samples, with the
exception of the most diluted one, for which $\Dinf = 0.54 h_0$.

In order to explain the volume fraction dependence of the solid
friction characteristic length, we propose a simple argument based
on the assumption that the solid friction stress between the wall
and the gel is proportional to $N$, the number of particles per unit
area in contact with the wall. During settling the gel is
compressed; as a result some particles are pushed against the wall.
The lower $\varphi_0$, the weaker the gel and the larger the number
of particles pushed on the wall, thus leading to enhanced solid
friction and smaller values of $L$. This argument may seem in
contradiction with Amonton's law which sates that the frictional
force $F$ is proportional to the load $P$ and is independent of the
area $A$ of the surfaces in contact \cite{Dowson1979}. However, as
first observed by Coulomb, in the presence of adhesion forces $F$ is
a linear function of both load and contact area: $F= \mu P + cA$. As
explained by Ringlein and Robbins \cite{RingleinAJP2004}, if the
real contact area $A$ is proportional to the load, as it is the case
for elastic solids, Amonton's law follows automatically: $F=\mu_{s}
P$ with $\mu_{s} =\mu + kc$. So, our assumption that  the the solid
friction is proportional to the number of solid particles per unit
area in contact with the wall will hold if the following conditions
are fulfilled: (i) the friction force between the gel and the
container is mainly dominated by the adhesion of solid particles to
the wall ($k c \gg \mu $); (ii) adhesive forces are negligible in
determining the flattening of contact between the gel and the wall
(the compaction of the gel is dominated by the gravity field). Both
conditions seem realistic. When the gel  is formed, the aggregates
can be described as spherical blobs of volume $V\sim\xi^{3}$, with a
negligible number of solid bonds in direct contact with the wall
(zero initial friction). As the equilibrium compaction of the gel is
reached, the relative variation of the blob volume, which we assume
to be isotropic for simplicity, can be approximated by the first
term on the r.h.s. of Eq.~(\ref{Eq:DeltaapproxIIb}): $\Delta V / V
\sim \Dinf/h_{0} \sim L/\lambda \sim \Delta \xi/\xi$, where the last
relation applies to small deformations. For a fractal blob, the
number of particles scales with blob size as $n \sim \xi^{d_f}$;
therefore, the number of solid bonds per blob which will touch the
wall due to a change in size $\Delta \xi$ is $\Delta n \sim
\xi^{d_f-1} \Delta \xi \sim \xi^{d_f}L/\lambda $. The number of new
contacts per unit area is $N \sim \Delta n/\xi^2 \sim
\xi^{d_f-2}L/\lambda$; using the scaling $L/\lambda \sim
\varphi_0^{-1.5}$ obtained experimentally and $\xi \sim
\varphi_0^{1/(d_f-3)}$, for $d_f = 1.9$ one finally finds
\begin{equation}
L \sim N^{-1} \sim \varphi_{0}^{1.4 \pm 0.4} \, ,
\label{Eq:scalingL}
\end{equation}
where the uncertainty in the exponent has been calculated assuming
an uncertainty of 0.05 in the fractal dimension. This expression
captures correctly the \emph{decrease} of the solid friction between
the wall and the gel as its volume fraction \emph{increases}, as
observed experimentally, although the exponent in
Eq.~(\ref{Eq:scalingL}) is lower than the measured one. In view of
the large experimental uncertainties in the scaling of $L$ (see
Fig.~\ref{Fig:ellelambdavsphi}) and the high sensitivity to the
exact value of $d_f$ of the exponent, the agreement seems however
reasonable.

As a final remak, we note that the model proposed in this work may
explain some features in previously published data on the settling
of strong gels. In Fig. 1 of \rref{SenisPRE1997}, Senis and Allain
plot $\Omega = 1 - \Dinf/h_0$ as a function of volume fraction. A
power law behavior is observed, except for $\varphi_0 \gtrsim 0.01$,
for which $\Omega$ tends to saturate to one, since for
high-volume-fraction, stiff gels $\Dinf << h_0$. One may wonder
whether our model may apply to these concentrated samples. Indeed,
we find that our data plotted as $\Omega$ $vs$ $\varphi_0$ fit on
the high volume fraction portion of the curve shown by Senis and
Allain (except for an irrelevant shift factor in volume fraction,
due to the difference in the physico-chemistry of the systems). This
strongly suggests that for $\varphi_0 \gtrsim 0.01$ the gels of
\rref{SenisPRE1997} retain their integrity and can be described by
our model. Furthermore, one can see in Fig. 2 of \rref{SenisPRE1997}
that for $\varphi_0 \ge 0.004$ and $h_0 \gtrsim 40$ mm $\Omega
\approx \mathrm{constant}$ similarly to our samples, implying that
$\Dinf \propto h_0$ and confirming that the model proposed here
applies also to the most concentrated samples of Senis and Allain.
Similarly, we wonder whether some of the data by Manley and
coworkers \cite{ManleyPRL2005Sedimentation} may fit in the framework
of our model. We recall that their system is very close to ours, the
main difference being the salt that is used to induce aggregation
($\mathrm{MgCl}_2$ at 20 mM rather than NaCl at 2 M) and the range
of volume fractions investigated ($0.005 \le \varphi_0 \le 0.08$ in
\cite{ManleyPRL2005Sedimentation}, as opposed to $0.0025 \le
\varphi_0 \le 0.01$). The fact that most of their data at large
volume fraction can be reproduced neglecting solid friction is
consistent with our finding that $L$ decreases with increasing
$\varphi_0$, making solid friction less important for more
concentrated gels. However, for the less concentrated gel and the
tallest samples the $h_0$ dependence of $\tau$ shown in Fig. 3 of
\rref{ManleyPRL2005Sedimentation} clearly departs from the quadratic
law predicted in the absence of solid friction. If a similar
deviation was observed also for $\Dinf~vs~h_0$, this would suggest
that for their most diluted and tallest samples solid friction does
indeed play a role.

\section{Conclusions}
\label{sec:con}

We have investigated the settling of rather diluted colloidal gels
made of strongly bound silica particles, focussing on the small
deformation regime where the network integrity is preserved.
Contrary to previous works, we find that the asymptotic settling is
incompatible with a model based only on a linear elastic response to
the gravitational stress. By contrast, we are able to model our
results by adding a solid friction term to the balance of forces
acting on the gel. We find that solid friction plays a crucial role
as long as the associated screening length $L$ is smaller than both
the gel height and the characteristic length scale $\lambda$ for
elastic deformation. Therefore, solid friction should matter for
tall gels at moderate to small volume fractions. By contrast, solid
friction is negligible when $L$ is comparable to or larger than the
gel height, regardless of the magnitude of $\lambda$. Additionally,
we have shown that, due to the fractal morphology of the aggregates
composing the gel, the contribution of solid friction becomes more
important as the gel volume fraction decreases, a somehow surprising
result.

Our work provides a more general framework for understanding and
modeling the asymptotic behavior of settling gels and could easily
be extended to other viscoelastic media. Future investigations
should address the full kinetics of sedimentation, where we expect
solid friction to play again a role, together with viscous
dissipation and elastic forces.

\ack

We acknowledge financial support from CNES (grants no. 02/4800000063
and 03/4800000123), and the European Community through the
``Softcomp'' Network of Excellence and the Marie Curie Research and
Training Network ``Arrested Matter''(grant no. MRTN-CT-2003-504712).
L. C. thanks the Institut Universitaire de France for supporting his
research.

\end{document}